\documentclass[ALICE,manyauthors]{cernphprep}
\usepackage[comma,square,numbers,sort&compress]{natbib}
\usepackage{hyperref}
\usepackage{lineno}
\usepackage{xspace}
\usepackage{xcolor}
\usepackage[version=4]{mhchem}
\usepackage{pdflscape}
\usepackage{array}
\usepackage{multirow}
\usepackage{makecell}
\usepackage[T1]{fontenc}
\usepackage{orcidlink} 

\begin{document}
%

\newcommand{\pp}           {pp\xspace}
\newcommand{\ppbar}        {\mbox{$\mathrm {p\overline{p}}$}\xspace}
\newcommand{\XeXe}         {\mbox{Xe--Xe}\xspace}
\newcommand{\PbPb}         {\mbox{Pb--Pb}\xspace}
\newcommand{\pA}           {\mbox{pA}\xspace}
\newcommand{\pPb}          {\mbox{p--Pb}\xspace}
\newcommand{\AuAu}         {\mbox{Au--Au}\xspace}
\newcommand{\dAu}          {\mbox{d--Au}\xspace}

\newcommand{\s}            {\ensuremath{\sqrt{s}}\xspace}
\newcommand{\snn}          {\ensuremath{\sqrt{s_{\mathrm{NN}}}}\xspace}
\newcommand{\pt}           {\ensuremath{p_{\rm T}}\xspace}
\newcommand{\pts}          {\ensuremath{p_{\rm T}^{\rm s}}\xspace}
\newcommand{\ptj}          {\ensuremath{p_{\rm T}^{\rm jet}}\xspace}
\newcommand{\meanpt}       {\ensuremath{\langle p_{\mathrm{T}}\rangle}\xspace}
\newcommand{\meanz}        {\ensuremath{\langle z \rangle }\xspace}
\newcommand{\meanpts}       {\ensuremath{\langle p_{\rm T}^{\rm s}\rangle}\xspace}
\newcommand{\ycms}         {\ensuremath{y_{\rm CMS}}\xspace}
\newcommand{\ylab}         {\ensuremath{y_{\rm lab}}\xspace}
\newcommand{\etarange}[1]  {\mbox{$\left | \eta \right |~<~#1$}}
\newcommand{\yrange}[1]    {\mbox{$\left | y \right |~<~#1$}}
\newcommand{\dndy}         {\ensuremath{\mathrm{d}N_\mathrm{ch}/\mathrm{d}y}\xspace}
\newcommand{\dndeta}       {\ensuremath{\mathrm{d}N_\mathrm{ch}/\mathrm{d}\eta}\xspace}
\newcommand{\avdndeta}     {\ensuremath{\langle\dndeta\rangle}\xspace}
\newcommand{\dNdy}         {\ensuremath{\mathrm{d}N_\mathrm{ch}/\mathrm{d}y}\xspace}
\newcommand{\Npart}        {\ensuremath{N_\mathrm{part}}\xspace}
\newcommand{\Ncoll}        {\ensuremath{N_\mathrm{coll}}\xspace}
\newcommand{\dEdx}         {\ensuremath{\textrm{d}E/\textrm{d}x}\xspace}
\newcommand{\RpPb}         {\ensuremath{R_{\rm pPb}}\xspace}
\newcommand{\sumpt}  {\ensuremath{\sum {\pt}}} 
\newcommand{\sumptc} {\ensuremath{\sum {\pt}^{\rm assoc}}} 
\newcommand{\sumptns} {\ensuremath{\sum^{\rm NS} {\pt}}} 

\newcommand{\nineH}        {$\sqrt{s}~=~0.9$~Te\kern-.1emV\xspace}
\newcommand{\seven}        {$\sqrt{s}~=~7$~Te\kern-.1emV\xspace}
\newcommand{\twoH}         {$\sqrt{s}~=~0.2$~Te\kern-.1emV\xspace}
\newcommand{\twosevensix}  {$\sqrt{s}~=~2.76$~Te\kern-.1emV\xspace}
\newcommand{\five}         {$\sqrt{s}~=~5.02$~Te\kern-.1emV\xspace}
\newcommand{\twosevensixnn}{$\sqrt{s_{\mathrm{NN}}}~=~2.76$~Te\kern-.1emV\xspace}
\newcommand{\fivenn}       {$\sqrt{s_{\mathrm{NN}}}~=~5.02$~Te\kern-.1emV\xspace}
\newcommand{\LT}           {L{\'e}vy-Tsallis\xspace}
\newcommand{\cmc}          {cm/$c$\xspace}
\newcommand{\GeVc}         {Ge\kern-.1emV/$c$\xspace}
\newcommand{\MeVc}         {Me\kern-.1emV/$c$\xspace}
\newcommand{\TeV}          {Te\kern-.1emV\xspace}
\newcommand{\GeV}          {Ge\kern-.1emV\xspace}
\newcommand{\MeV}          {Me\kern-.1emV\xspace}
\newcommand{\GeVmass}      {Ge\kern-.2emV/$c^2$\xspace}
\newcommand{\MeVmass}      {Me\kern-.2emV/$c^2$\xspace}
\newcommand{\lumi}         {\ensuremath{\mathcal{L}}\xspace}

\newcommand{\ITS}          {\rm{ITS}\xspace}
\newcommand{\TOF}          {\rm{TOF}\xspace}
\newcommand{\ZDC}          {\rm{ZDC}\xspace}
\newcommand{\ZDCs}         {\rm{ZDCs}\xspace}
\newcommand{\ZNA}          {\rm{ZNA}\xspace}
\newcommand{\ZNC}          {\rm{ZNC}\xspace}
\newcommand{\SPD}          {\rm{SPD}\xspace}
\newcommand{\SDD}          {\rm{SDD}\xspace}
\newcommand{\SSD}          {\rm{SSD}\xspace}
\newcommand{\TPC}          {\rm{TPC}\xspace}
\newcommand{\TRD}          {\rm{TRD}\xspace}
\newcommand{\VZERO}        {\rm{V0}\xspace}
\newcommand{\VZEROA}       {\rm{V0A}\xspace}
\newcommand{\VZEROC}       {\rm{V0C}\xspace}
\newcommand{\Vdecay} 	   {\ensuremath{V^{0}}\xspace}

\newcommand{\ee}           {\ensuremath{e^{+}e^{-}}} 
\newcommand{\pip}          {\ensuremath{\pi^{+}}\xspace}
\newcommand{\pim}          {\ensuremath{\pi^{-}}\xspace}
\newcommand{\kap}          {\ensuremath{\rm{K}^{+}}\xspace}
\newcommand{\kam}          {\ensuremath{\rm{K}^{-}}\xspace}
\newcommand{\pbar}         {\ensuremath{\rm\overline{p}}\xspace}
\newcommand{\kzero}        {\ensuremath{{\rm K}^{0}_{\rm{S}}}\xspace}
\newcommand{\lmb}          {\ensuremath{\Lambda}\xspace}
\newcommand{\almb}         {\ensuremath{\overline{\Lambda}}\xspace}
\newcommand{\Om}           {\ensuremath{\Omega^-}\xspace}
\newcommand{\Mo}           {\ensuremath{\overline{\Omega}^+}\xspace}
\newcommand{\X}            {\ensuremath{\Xi^-}\xspace}
\newcommand{\Ix}           {\ensuremath{\overline{\Xi}^+}\xspace}
\newcommand{\Xis}          {\ensuremath{\Xi^{\pm}}\xspace}
\newcommand{\Oms}          {\ensuremath{\Omega^{\pm}}\xspace}
\newcommand{\degree}       {\ensuremath{^{\rm o}}\xspace}


\newcommand{\sNN}{\ensuremath{\snn}}
\newcommand{\RAA}{\ensuremath{R_{\rm AA}}\xspace}
\newcommand{\TAA}{\ensuremath{T_{\rm AA}}\xspace}

\newcommand{\RpPbjch} {\ensuremath{\RpPb^{\rm ch~jet}}\xspace}

\newcommand{\pTtrk}{\ensuremath{p_{\rm T,track}}\xspace}
\newcommand{\pTj}{\ensuremath{p_{\rm T,jet}}\xspace}
\newcommand{\pTjch}{\ensuremath{\pTj^{\rm ch}}\xspace}
\newcommand{\pTjchraw}{\ensuremath{\pTj^{\rm ch~raw}}\xspace}
\newcommand{\pTjlead}{\ensuremath{\pTj^{\rm ch~lead}}\xspace}
\newcommand{\pTjdet}{\ensuremath{\pTj^{\rm ch~det}}\xspace}
\newcommand{\pTjtruth}{\ensuremath{\pTj^{\rm ch~truth}}\xspace}

\newcommand{\etaj}{\ensuremath{\eta_{\rm jet}}\xspace}
\newcommand{\Ajet}{\ensuremath{A_{\rm jet}}\xspace}
\newcommand{\rhobkg}{\ensuremath{\rho_{\rm bkg}}\xspace}
\newcommand{\rhoch}{\ensuremath{\rho_{\rm ch}}\xspace}
\newcommand{\kT}{\ensuremath{k_{\rm T}}\xspace}
\newcommand{\akT}{anti-$\kT$\xspace}
\newcommand{\sigRC}{\ensuremath{\sigma^{\rm RC}}\xspace}
\newcommand{\thirteen} {$\sqrt{s}~=~13$~Te\kern-.1emV\xspace}

\newcommand{\jpsi}   {\rm J/$\psi$}
\newcommand{\psip}   {$\psi^\prime$}
\newcommand{\jpsiDY} {\rm J/$\psi$\,/\,DY}
\newcommand{\dd}     {\mathrm{d}}
\newcommand{\chic}   {$\chi_{\rm c}$}
\newcommand{\ezdc}   {$E_{\rm ZDC}$}

\newcommand{\mup}{\ensuremath{\mu^{+}}}
\newcommand{\mum}{\ensuremath{\mu^{-}}}
\newcommand{\mupm}{\ensuremath{\mu^{\pm}}}
\newcommand{\Jpsi}{\ensuremath{{\rm J/}\psi}}

\newcommand{\pipm}{\ensuremath{\pi^{\pm}}}
\newcommand{\Kpm}{\ensuremath{{\rm K}^{\pm}}}

\newcommand{\Vzero}{\ensuremath{{\rm V}^{0}}\xspace}
\newcommand{\Vzeros}{$\Vzero$s\xspace}

\newcommand{\Dzero}{\ensuremath{{\rm D}^{0}}}
\newcommand{\Bpm}{\ensuremath{{\rm B}^{\pm}}}

\newcommand{\cent}   [2] {$#1$--$#2\%$}
\newcommand{\slfrac} [2] {\left.#1\right/#2}

\newcommand{\abs}[1]{\ensuremath{\left|#1\right|}}
\newcommand{\avg}[1]{\ensuremath{\left\langle#1\right\rangle}}

\newcommand{\Geant}{\textsc{Geant $3$}\xspace}
\newcommand{\Pythia}{\textsc{Pythia}\xspace}
\newcommand{\Pyeight}{\textsc{Pythia $8$}\xspace}
\newcommand{\Pyeightthree}{\textsc{Pythia $8.3$}\xspace}
\newcommand{\Pysix}{\textsc{Pythia $6$}\xspace}
\newcommand{\Pdratio}{\Pythia-to-data\xspace}
\newcommand{\Pwgpy}{\textsc{Powheg+Pythia $8$}\xspace}
\newcommand{\Pwg}{\textsc{Powheg}\xspace}
\newcommand{\Pwgbox}{\textsc{Powheg~Box}\xspace}
\newcommand{\Scape}{\textsc{Jetscape}\xspace}

\newcommand{\CSec}[1]{Section~\ref{#1}}
\newcommand{\ssec}[1]{section~\ref{#1}}

\newcommand{\emp}{\textcolor{red}}
\newcommand{\mrk}{\textcolor{magenta}}

\begin{titlepage}
\PHyear{2026}       
\PHnumber{074}      
\PHdate{12 March}  

\title{Measurement of the transverse-momentum fraction of strange hadrons from jet-like correlation structures in pp collisions at $\sqrt{s}={\bf 13}$ TeV}
\ShortTitle{Transverse-momentum fraction of strange particles from jet-like correlations}   

\Collaboration{ALICE Collaboration\thanks{See Appendix~\ref{app:collab} for the list of collaboration members}}
\ShortAuthor{ALICE Collaboration} 

\begin{abstract}
The first measurements of the average transverse-momentum fraction (\meanz) as a function of transverse momentum (\pt) for strange baryons (\lmb and \almb) and strange mesons (\kzero), produced in mini-jets defined through angular correlations in \pp collisions at \thirteen, are reported by the ALICE Collaboration at the LHC.
The observable is obtained using a novel method, where the angular correlation between the strange hadrons and inclusive charged hadrons is weighted by the \pt of correlated particles at small angular distance.
As a function of strange particles' \pt, the results reveal a flat trend for strange mesons and a decreasing trend for strange baryons in the measured \pt region, indicating distinct hadronization mechanisms for \kzero and \lmb~(\almb).
The measurements are compared to Monte Carlo models, namely \Pyeight (with both Monash and Color Rope tunes) and the AMPT (A Multi-Phase Transport) model with string melting.
None of these models provides a satisfactory description of the $\meanz$ distributions at low and intermediate \pt.
\end{abstract}

\end{titlepage}

\setcounter{page}{2} 


\section{Introduction}
\label{sec:Intro}
Hadronic collisions at the LHC (\pp, \pPb, and \PbPb) provide the opportunity to study particle production across a wide range of charged-particle multiplicities and corresponding initial energy densities~\cite{ALICE:2018pal, ALICE:2019dfi, ALICE:2014xsp, ALICE:2012aqc, ALICE:2012ovd, ALICE:2022wpn}.
One of the most intriguing discoveries at the LHC is that phenomena traditionally associated with heavy-ion collisions, and attributed to the formation of quark--gluon plasma (QGP), exhibit a continuous onset as a function of multiplicity across all collision systems. These features include collective fluid-like behavior~\cite{ALICE:2024vzv, CMS:2010ifv, ATLAS:2015hzw}, strangeness enhancement, and baryon-to-meson yield enhancement in the intermediate \pt range~\cite{ALICE:2016fzo, ALICE:2013wgn, ALICE:2019avo}.

In the so-called ``small systems", such as \pp and \pPb collisions, the size of the interaction region is approximately one femtometer. 
At LHC energies, partonic systems produced in multiparton interactions (MPIs) are expected to overlap within this small region.
Given that the range of the strong force is of comparable size, interactions among strings produced nearby at early times may induce effects on the final-state kinematics and hadronization beyond those expected from a simple incoherent superposition of MPIs. 
In \PbPb collisions, final-state interactions are well established and give rise to strong collective effects.

The presence of collective effects in small systems raises the intriguing possibility that droplets of QGP may already be forming in such systems.
Furthermore, understanding how MPI-driven effects manifest themselves in small systems can shed light on collective phenomena observed in Pb–Pb collisions.
Addressing these issues is important for developing a common understanding of particle production mechanisms across small and large collision systems, and for identifying the conditions that give rise to QGP-like behavior.

The enhancement of strange baryon-to-meson hadron yield ratios ($\Lambda / \kzero$, $\Xi / \kzero$, $\Omega / \kzero$) in the intermediate \pt\ region, observed across all collision systems and progressively increasing with multiplicity, has been attributed to the interplay between radial flow and parton recombination~\cite{Muller:2012zq, Fries:2003vb, Bozek:2011gq}. 
To further investigate the origin of this effect, ALICE has measured these ratios within charged-particle jets with \pt above 10 \GeVc and for particles in the underlying event of these jets~\cite{ALICE:2021vxl, ALICE:2022ecr}. 
The results show that, within the precision of the measurements, the enhancement is small or absent within jets, while in the underlying event, it matches the inclusive results.
However, a limitation of this approach is that the production of a particle outside a jet with a minimum energy threshold does not necessarily indicate it was not produced in a hard-scattering process.
For example, the increase of the ratios between particle yields measured at $\sqrt{s}=13$ TeV and 7 TeV with rising \pt suggests that hard scatterings begin to dominate particle production already for $\pt > 2$ \GeVc~\cite{ALICE:2020jsh}.
This raises the need for a complementary observable to study particle production mechanisms. 
Assuming that particle production is predominantly governed by parton fragmentation, the fraction of the parton's transverse momentum carried by the final-state particle (\pt/\ptj) provides complementary insight into the fragmentation process and is therefore a more informative observable than the hadron \pt alone.

In the low- and intermediate-\pt region (\pt < 5 GeV/c), the distinction between jets and soft bulk particle production becomes ambiguous, effectively precluding reliable event-by-event jet reconstruction.
Therefore, the present analysis does not rely on associating strange particles with reconstructed jet cones.
Instead, the average momentum fraction (\meanz) of \lmb~(\almb) and \kzero hadrons in mini-jets is measured. 
The mini-jets are defined via $\Delta\varphi$–$\Delta\eta$ angular correlations between these strange hadrons and primary charged particles, where $\Delta\varphi$ and $\Delta\eta$ denote the differences in azimuthal angle and pseudorapidity, respectively.
The angular differential distribution is weighted by the \pt of the associated particles in order to estimate the \pt of the originating partons.

Jet-like structures manifest themselves as a short-range near-side peak centered at $\Delta\varphi = \Delta\eta = 0$ in the angular correlation distribution.
After normalizing the \pt-weighted angular correlation distribution by the number of trigger particles and subtracting the uncorrelated background, the integral below the short-range near-side peak measures the summed \pt correlated with the strange particles (\sumptns).
A proxy for the original parton \pt can then be obtained by adding the \pt of the trigger particle to \sumptns. 
This approach provides essential complementary information for the understanding of particle production dynamics and the role of parton fragmentation in the intermediate \pt region.

\section{Experimental setup and data samples}
\label{sec:ExpSetupAndDataSample}
The ALICE detector~\cite{ALICE:2008ngc} has been optimized for the study of heavy-ion collisions at the LHC.
The main objective of ALICE is to explore the physics of strongly-interacting matter in proton-proton collisions and the QGP produced in the extreme conditions of temperature and density created in nucleus-nucleus collisions.
The ALICE detector and its performance during Run 2 of the LHC are described in Refs.~\cite{ALICE:2014sbx, ALICE:2013axi}.
Its setup consists of central-barrel detectors covering the pseudorapidity range \etarange{0.9}, forward detectors, and the magnet systems.

The central-barrel detectors -- Inner Tracking System (\ITS), Time Projection Chamber (\TPC), Transition Radiation Detector (\TRD), Time Of Flight (\TOF), Photon Spectrometer (PHOS), Electromagnetic Calorimeter (EMCal), and High Momentum Particle Identification Detector (HMPID) -- are surrounded by the L3 solenoid, which provides a magnetic field of up to $0.5$~T in strength along the beam direction.
These detectors are optimized for tracking and particle identification.
The Photon Multiplicity Detector (PMD), Forward Multiplicity Detector (FMD), \VZERO, T0, and Zero Degree Calorimeter (\ZDC) -- the forward detectors  -- provide triggering and information on global event properties. 
The measurements presented in this paper use the information from \VZERO~\cite{ALICE:2013axi}, \ITS~\cite{ALICE:2013nwm}, \TPC~\cite{ALICE:2000jwd}, and \TOF~\cite{ALICE:2000xcm, Strazzi:2025reb}, which are described in more detail in the following paragraphs.

The \VZERO is used to provide event triggering information. 
It is composed of two scintillator arrays, \VZEROA and \VZEROC, placed around the beam pipe on each side of the interaction point, covering the pseudorapidity intervals $2.8~<~\eta~<~5.1$ and $-3.7~<~\eta~<~-1.7$, respectively.

The \ITS is the detector closest to the interaction region.
The \ITS used in Run 2 was made of six cylindrical layers of silicon detectors.
The two innermost layers comprise the Silicon Pixel Detector (\SPD), which mainly provides hit position information for vertexing and tracking.
The intermediate two layers form the Silicon Drift Detectors (\SDD), and the outermost two layers form the Silicon Strip Detectors (\SSD).
They mainly provide tracking information and particle identification via the measurement of ionization energy loss.

The \TPC is the main tracking detector of ALICE.
It is a cylindrical detector with a volume of about 90~${\rm m}^{3}$, designed to be filled with a gas mixture of 90\% \ce{Ne} and 10\% \ce{CO2}. 
By measuring ionization energy loss, the \TPC is also used for particle identification.

The \TOF, placed outside of the \TPC, is made of Multigap Resistive Plate Chambers (MRPC).
Complementary to the \ITS and the \TPC, the \TOF provides particle identification at the intermediate momentum range (from 0.5 \GeVc to 3--4 \GeVc)~\cite{Strazzi:2025reb} via time-of-flight measurements.

The present analysis is performed with \pp collision data at $\s =$~13 TeV, which were collected by ALICE during the LHC Run 2 from 2016 to 2018.
The events used for the analysis are those satisfying the minimum 
bias (MB) trigger condition requiring at least one hit in both 
\VZEROA and \VZEROC.
Pileup events are rejected using vertex and tracking information.
The events are accepted only if the reconstructed primary 
vertex position in the beam direction is located in the range $\pm 10$~cm from the ALICE nominal interaction point.
Approximately $1.7 \times 10^9$ events pass these selection criteria.

Monte Carlo (MC) data samples have been generated using  
\Pyeight with the Monash tune~\cite{Sjostrand:2014zea, Skands:2014pea} as the primary event generator for pp 
collisions. The detector response is simulated by transporting the
primary events through a model of the ALICE detector using 
\Geant~\cite{Brun:1994aa}.
These MC samples are used to evaluate the tracking efficiency and the secondary contamination contribution.

\section{Analysis procedure}
\label{sec:AnaApproach}
The goal of the present analysis is to study an observable that can approximate the transverse-momentum fraction $z$ of strange particles in low-energy jets. 
For reconstructed jets, $z$ is defined as $z~=\pts/\ptj$, where $\pts$ is the transverse momentum of the strange particle and $\ptj$ the transverse momentum of the jet. 
Very low-energy jets ($\ptj \ll $~5~\GeVc) cannot be reconstructed event-by-event over the fluctuating background of the underlying event. 
This is particularly true in high-multiplicity pp collisions. 
An alternative method to assess the fragmentation properties of low-energy jets is the study of $\Delta\varphi - \Delta \eta$ angular correlations between particles, given that jet-like correlations produce a distinct peak in the correlation function centered at $\Delta\varphi = \Delta \eta = 0$, the so-called short-range near-side peak.
In the angular correlation study, $\Delta\varphi$ and $\Delta\eta$ are defined as the azimuthal angle and pseudorapidity difference between the trigger and the associated particles, respectively.
In this measurement, the strange particles are used as trigger particles, and all primary charged particles are used as associated particles.
Primary charged particles are defined as charged particles with a mean proper lifetime $\tau$ larger than 1 \cmc, which are produced either promptly at the primary interaction vertex or from decays of particles with $\tau$~$<$~1~\cmc restricted to decay chains leading to the interaction~\cite{ALICE:2017hcy}.
For the present analysis, the long-lived charged-strange baryons ($\Sigma^{\pm}$, $\Xi^{-}$, $\Omega^{-}$, and their antiparticles) are excluded from the primary charged-particle definition in order to avoid biases, as their reconstruction requires dedicated topological selections beyond those used for primary charged particles.

\subsection{Particle reconstruction}
\label{sec:ParticleRec}
The strange particles, \kzero, \lmb, and \almb, are reconstructed through their decay daughters in the central barrel region via the $\kzero \to \pip + \pim~$, $\lmb \to {\rm p} + \pim~$, and $\almb \to \overline{{\rm p}} + \pip$ decay channels, with the branching ratios of $69.20\pm0.05\%$, $64.1\pm0.5\%$, and $64.1\pm0.5\%$, respectively~\cite{ParticleDataGroup:2024cfk}.
The reconstruction is performed in the kinematic range 0.6 $<$ \pt $<$ 20.0 \GeVc and \etarange{0.75}.
They are selected based on their characteristic V-shaped decay topology.
Being electrically neutral, these strange particles are also referred to as \Vzero particles.
The selection criteria include the \Vzero decay radius in the transverse plane, the distance of the closest approach (DCA) between the daughter tracks, the cosine of the pointing angle (CPA), which is the angle between the line connecting the PV to the decay vertex and the direction of the momentum vector.
Additional selection criteria are applied to the daughter tracks of the \Vzero, including pseudorapidity requirements, the DCA of each daughter to the primary vertex (PV), and PID and track-quality selections based on the \TPC information.
These criteria follow those presented in an earlier ALICE publication~\cite{ALICE:2022ecr} and are summarized in Table~\ref{table: V0Selection}.
To suppress misidentification between \kzero and \lmb~(\almb) candidates, the invariant mass is also computed under the competing mass hypothesis, and a selection criterion is applied accordingly.
In addition, in order to reject strange particle candidates from out-of-bunch pileup, at least one daughter track is required to satisfy the ITS refit condition, exploiting the superior time resolution of the \ITS.

\begin{table}[th]
    \caption{$\Vzero$ candidate selection criteria.}
    \label{table: V0Selection}
    \centering
    \begin{tabular}{ c c c } 
        \hline
        \rule{0pt}{2.5ex}
        \textbf{\Vzero selection} & $\kzero$ & $\lmb$ ($\almb$) \\ [0.5ex]
        \hline
        \abs{\eta_{\rm \Vzero}}                          & $<0.75$      & $<0.75$    \\ [0.2ex]
        $\Vzero$ 2D decay radius (cm)  & $>0.5 $      & $>0.5 $    \\ [0.2ex]
        Proper lifetime (\cmc)                             & $<20$         & $<30$       \\ [0.2ex]
        DCA between $\Vzero$ daughter tracks (N$\sigma$) & $<1$          & $<1$        \\ [0.2ex]
        CPA of $\Vzero$                                  & $>0.97$       & $>0.995$     \\ [0.2ex]
        Competing mass (\GeVmass)                        & $>0.005$     & $>0.010$   \\ [0.2ex]
        \hline
        \multicolumn{3}{c}{\rule{0pt}{2.5ex}\textbf{\Vzero daughters selection}} \\ [0.5ex]
        \hline
        $\abs{\eta_{\rm trk}}$                 & \multicolumn{2}{c}{$<0.8$} \\ [0.2ex]
        \Vzero daughter track DCA to PV (cm)   & \multicolumn{2}{c}{$>0.06$} \\ [0.2ex]
        TPC \dEdx (N$\sigma$)                  & \multicolumn{2}{c}{$<5$} \\ [0.2ex]
        TPC refit flag                         & \multicolumn{2}{c}{true} \\ [0.2ex]
        Number of TPC crossed rows             & \multicolumn{2}{c}{$>70$} \\ [0.2ex]
        TPC crossed rows / findable ratio      & \multicolumn{2}{c}{$>0.8$} \\ [0.2ex]
        \hline
    \end{tabular}
\end{table}

Primary charged particles are reconstructed in the kinematic range $0.2 < \pt < 20$~\GeVc and \etarange{0.8} using the \ITS and the \TPC.
Contamination from secondary particles is reduced by applying the selection criteria described in Ref.~\cite{ALICE:2021nvv}.
Selected primary charged tracks are required to have at least 70 crossed pad rows in the \TPC (out of a maximum of 159), and a ratio of crossed rows to the number of findable clusters of at least 0.8.
Only tracks with a fraction of shared clusters with other tracks smaller than 0.4 are accepted. 
The DCA to the primary vertex in the transverse plane ($xy$-plane) and the beam direction ($z$-direction) are required to be within 2.4 cm and 3.2 cm, respectively. 
The tracks are also required to have a good tracking fit quality in both the \TPC and the \ITS, characterized by goodness-of-fit values $\chi^{2}$ per cluster smaller than 4 and 36 in the \TPC and the \ITS, respectively.
Only tracks with at least one hit in the two innermost layers of the \ITS are selected.
Tracks originating from kink topologies, where a charged particle decays into another charged particle and a neutral particle, are rejected.

\subsection{\texorpdfstring{$\pt$}{pT}-weighted two-particle correlation function}
\label{sec:PtWeighted2PC}
A novel \pt-weighted two-particle correlation method is employed to obtain the average $z(\pts)$ of strange particles in any given \pts interval.
The corresponding correlation function is defined as:
\begin{equation}
C_{\pt\text{-weighted}}\left ( \Delta\eta, ~\Delta\varphi \right ) = 
\frac{1}{N_{\rm trig}}\frac{{\rm d^2}\sumpt}{{\rm d}\Delta\varphi{\rm d}\Delta\eta},
\label{eq:Wgt2PCCF}
\end{equation}
where $N_{\rm trig}$ is the number of trigger strange particles.
The term $\frac{{\rm d^2}\sumpt}{{\rm d}\Delta\varphi{\rm d}\Delta\eta}$ represents the summed \pt of all the associated particles in any given $\Delta\varphi - \Delta\eta $ interval.
Both $N_{\rm trig}$ and $\frac{{\rm d^2}\sumpt}{{\rm d}\Delta\varphi{\rm d}\Delta\eta}$ are summed over all selected events.

The correlation function is corrected for acceptance, efficiency, and uncorrelated background, which will be described in detail in the next section.
Finally, the summed \pt of the associated particles per trigger particle $\langle \sumptns \rangle$ is calculated by integrating the correlation function, and the average $\langle z(\pts)\rangle$ is computed as
\begin{equation}
	\centering
	\langle z (\pts ) \rangle  ~=~\frac{\pts}{\langle \sumptns \rangle + \pts}.
	\label{eq:zDef}
\end{equation}
As will be explained in the following paragraphs in more detail, \meanz is obtained as a function of the strange particle \pts by constructing the correlation function and applying the corrections independently in each \pts interval.
Due to the steeply falling trigger-particle \pts spectra and the finite width of the \pts intervals, especially at high transverse momenta, the \pts in Eq.~\ref{eq:zDef} is calculated as the average transverse momentum in each \pt interval instead of the value of the interval center.

It is important to emphasize the differences between the present approach and a standard jet analysis.
In the latter, jets are reconstructed event by event and the jet \pt is required to be either above a threshold value imposed by the jet reconstruction efficiency or within a specific range.
The $z$ is thus obtained event by event, and the resulting $z$ distribution might be affected by the jet \pt selection.
In contrast, the present analysis does not rely on explicit jet reconstruction and therefore does not impose any jet-\pt selection.
Since the jets are not reconstructed individually, an event-by-event determination of $z$ is not possible.
Instead, the observable $\langle z \rangle$ is obtained as an event-averaged quantity, calculated as \pts/$\langle$\ptj$\rangle$ in intervals of the trigger-particle \pts.

Let $D(z)$ denote the fragmentation function, the probability for a particle to carry a fraction $z$ of the originating parton momentum.
Assuming that the parton transverse momentum spectrum falls as $(1/\pt^{\rm parton})^n$, where typical values of $n$ are between 5 and 6 at high \pt~\cite{ALICE:2019dfi}.
At fixed trigger-particle transverse momentum $\pt$, and assuming that $z$ is measured on an event-by-event basis, the probability density of $z$ is proportional to $D(z) z^{n-2}$~\cite{Ellis:1976hun, Tannenbaum:2006ku}.
The factor $z^{n-2}$ biases the distribution towards larger values of $z$ compared to what would be obtained at fixed $\pt^{\rm parton}$.
This effect is known as the leading particle bias or trigger bias~\cite{Ellis:1976hun, Tannenbaum:2006ku}.
In the present analysis, one instead averages over $\sumptns$ which is proportional to $\pts / z$.
As a result, the weighting factor is reduced to $z^{n-3}$. This reduced trigger bias represents an advantage of the present method.

\subsection{Corrections}
\label{sec:Corrections}
The reconstructed strange particles consist of both real signal particles and combinatorial background (fake strange particles).
To obtain the correlation function for the signal particles, the contribution to the \pt-weighted correlation function from the combinatorial background is removed using the sideband subtraction method~\cite{ALICE:2019avo, ALICE:2021nvv}.
The invariant mass distributions of the strange particle candidates in each \pts interval are fitted with a Gaussian function for the signal combined with a linear function describing the combinatorial background.
Using the mean and variance ($\sigma$) of the Gaussian, the signal and the sideband regions are defined.
The signal region is defined as the $\pm 3 \sigma$ range centered at the mean value, while the sideband regions, used to estimate the combinatorial background, are located between 6$\sigma$ and 9$\sigma$ symmetric around the mean value.
The \pt-weighted correlation functions are constructed using triggers from the signal and sideband regions, respectively.
The sideband correlation distributions, which provide an estimation of the combinatorial background, are subtracted from the signal correlation distributions.

To account for tracking efficiency, a dedicated efficiency correction is applied to the associated particles.
The tracking efficiency ($\varepsilon$) is obtained from MC simulations and is defined as the ratio of the number of reconstructed particles to the number of generated particles satisfying the primary charged-particle definition (see Section~\ref{sec:Intro}).
In this work, the tracking efficiency increases from approximately 70\% at the lowest \pt, and reaches a plateau of 85\% above $\pt \approx 1.5$ \GeVc.
This correction is achieved by using $\varepsilon$ as another part of the weight when filling the \pt-weighted correlation function.
Not all charged tracks passing the selection criteria are primary tracks. 
The secondary contamination fraction ($f_{\rm sec}$) is defined as the ratio of secondary particles to the total number of particles passing the selection, and is also determined using MC simulations.
The estimated secondary fraction decreases from approximately 3.8\% at low \pt to below 1\% for $\pt > 1$ \GeVc.
To correct for both effects, an additional weight factor beyond the associated particle's \pt, namely $(1 - f_{\rm sec}) / \varepsilon$, is introduced when calculating the correlation functions. 

The corrected \pt-weighted correlation functions are still strongly affected by the limited and non-uniform acceptance of the detector.
In particular, the finite pseudorapidity acceptance leads to a characteristic triangular shape in the $\Delta\eta$ direction, which is a purely geometrical effect rather than a consequence of physical correlations.
To correct for this non-physical modification, the event-mixing technique is applied.
In this method, with the same selection criteria, the trigger particles and the associated particles are chosen from different events to construct the mixed-event correlation function, $M \left ( \Delta\eta, ~\Delta\varphi \right )~=~\frac{{\rm d^2}N_{\text{mix}}}{{\rm d}\Delta\varphi{\rm d}\Delta\eta}$, where $N_{\text{mix}}$ denotes the number of trigger-associate pairs in mixed events.
The events are mixed only if the difference in the $z$-vertex position is less than 2 cm.
Since trigger and associated particles come from different events, no physical correlation occurs; thus, the mixed-event distribution reflects only the detector effects.
For each trigger \pts interval, the \pt-weighted correlation function $C_{\pt\text{-weighted}}\left ( \Delta\eta, ~\Delta\varphi \right )$ and the mixed-event distribution $M\left ( \Delta\eta, ~\Delta\varphi \right )$ are constructed independently in different associated particle \pt intervals.
By dividing $C_{\pt\text{-weighted}}\left ( \Delta\eta, ~\Delta\varphi \right )$ and  $M\left ( \Delta\eta, ~\Delta\varphi \right )$, the corrected \pt-weighted correlation function is obtained as
\begin{equation}
    C^{\rm corrected}_{\pt\text{-weighted}}\left ( \Delta\eta, ~\Delta\varphi \right ) = 
    \frac{C_{\pt\text{-weighted}}\left ( \Delta\eta, ~\Delta\varphi \right )}{\alpha M\left ( \Delta\eta, ~\Delta\varphi \right )},
    \label{eq:MECorr}
\end{equation}
where $\alpha$ is the normalization factor for the mixed-event correlation function given by the inverse of the average along the $\Delta\varphi$-axis of bin contents for the bins where $\Delta\eta~=~0$.
This normalization is motivated by the fact that, at the limit of very small angular distance, the acceptance is expected to approach unity.

\begin{figure}[h]
    \begin{center}
    \includegraphics[width = 0.6\textwidth]{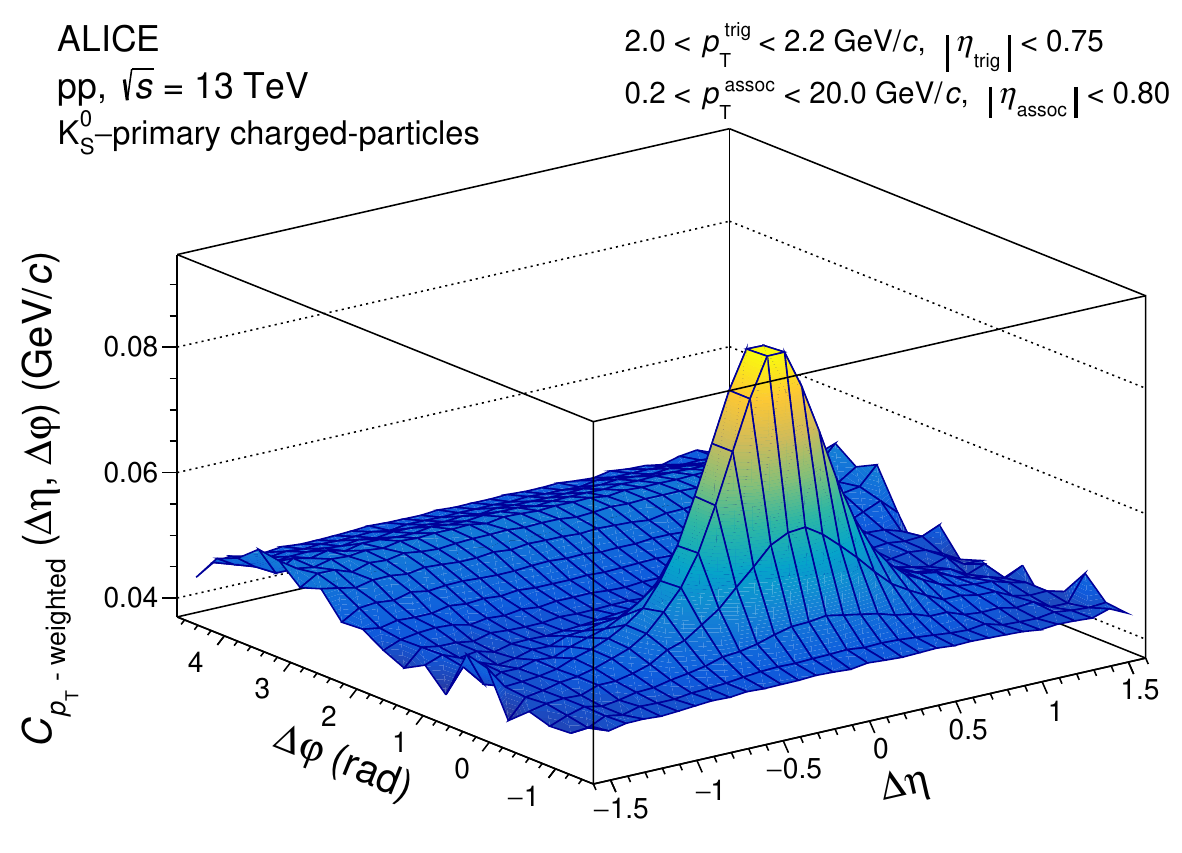}
    \includegraphics[width = 0.6\textwidth]{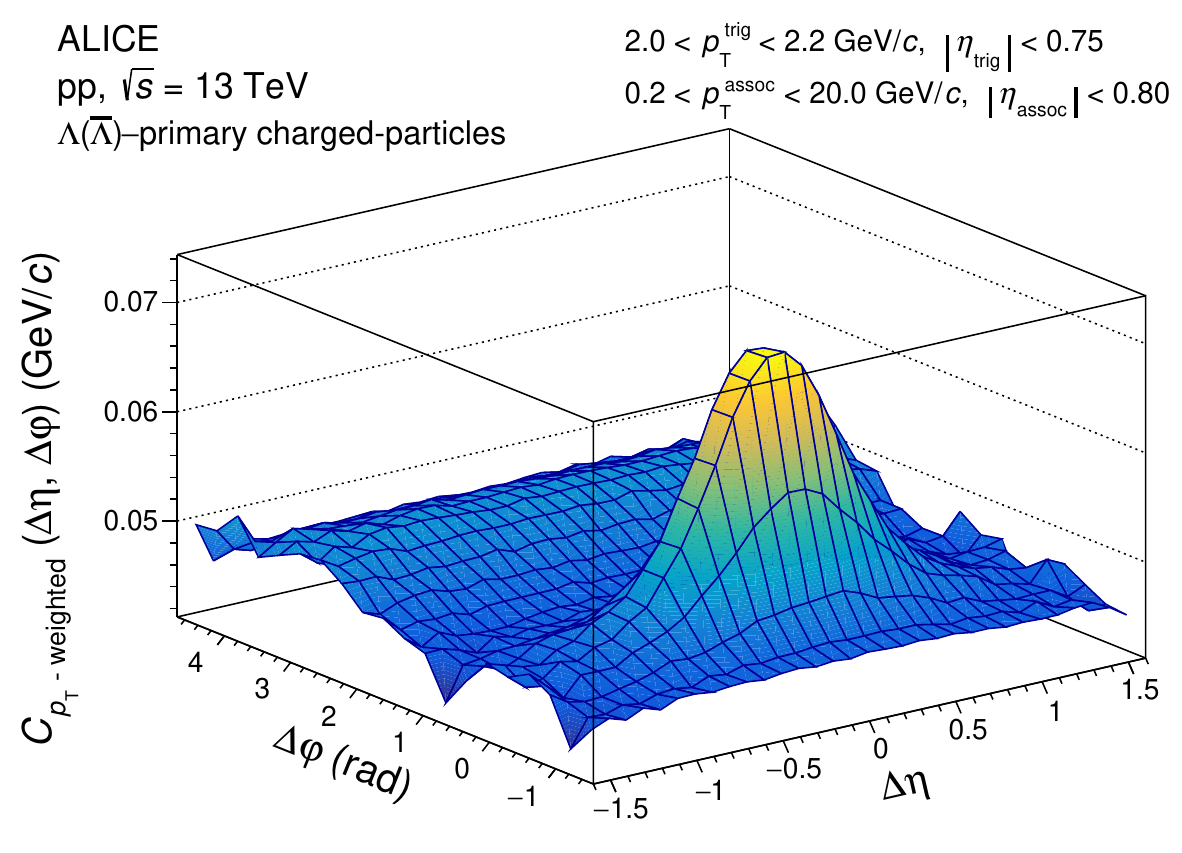}
    \end{center}
    \caption{Example of the fully corrected \pt-weighted correlation functions for \kzero--primary charged-particles correlation (top) and for \lmb~(\almb)--primary charged-particles correlation (bottom).}
    \label{fig:2DWgt2PCF}
\end{figure}

Finally, the corrected correlation functions from different associated particle \pt intervals are summed to obtain the final \pt-weighted correlation functions in each \pts interval.
Examples of the fully corrected \pt-weighted correlation functions for \kzero and \lmb~(\almb) are shown in Fig.~\ref{fig:2DWgt2PCF}.
The broader structure along $\Delta\eta$ for \lmb~(\almb) suggests a weaker collimation with respect to the jet axis compared to \kzero, possibly related to the different production mechanisms of baryons and mesons.

\begin{figure}[h]
    \begin{center}
    \includegraphics[width = 0.48\textwidth]{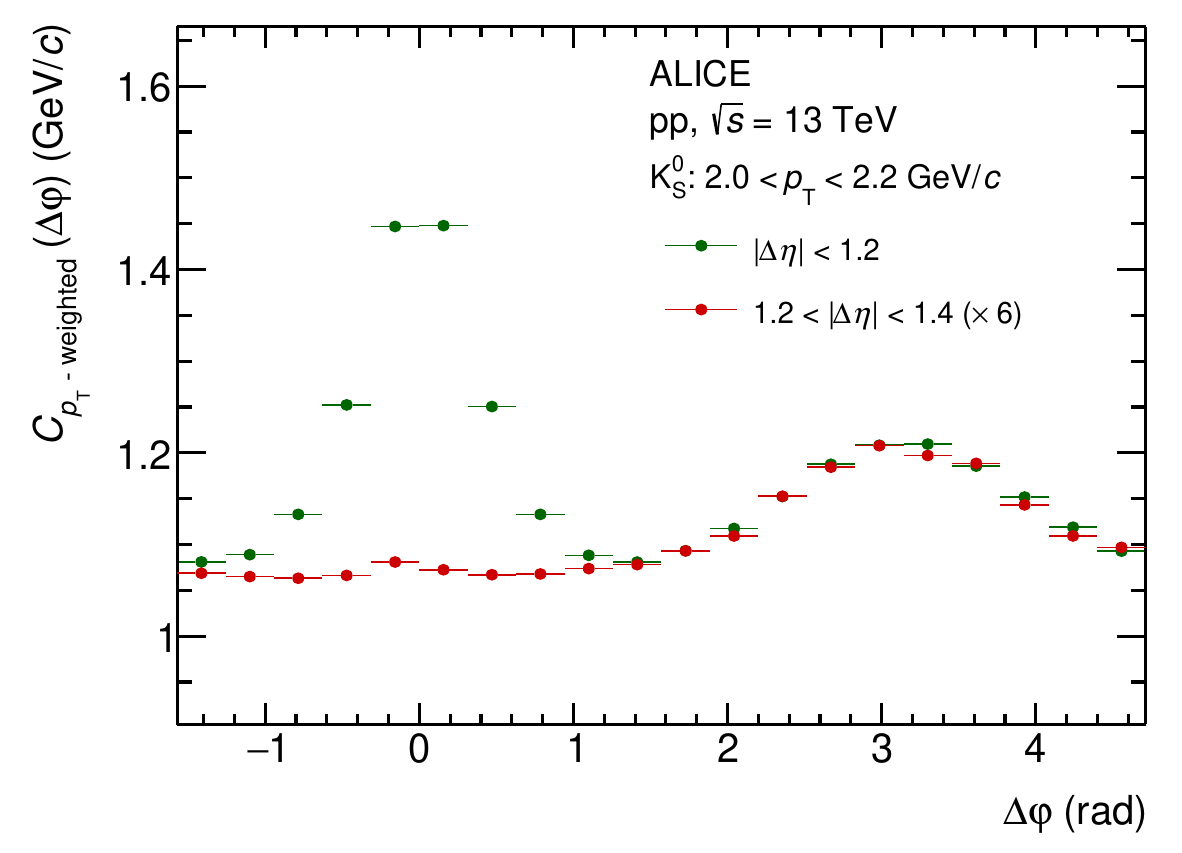}
    \includegraphics[width = 0.48\textwidth]{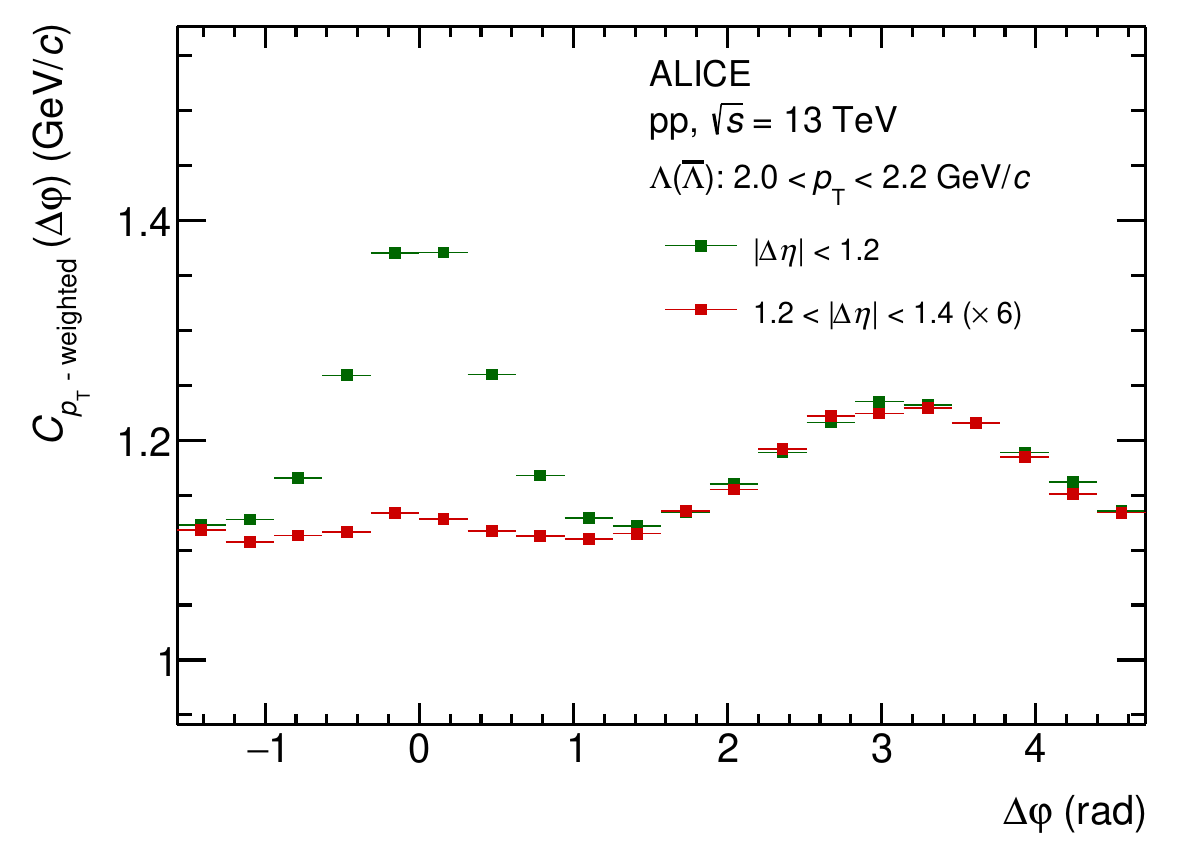}
    \end{center}
    \caption{Examples of $\Delta\varphi$ projections of the fully corrected \pt-weighted correlation functions for \kzero--primary charged-particles correlation (left) and for \lmb~(\almb)--primary charged-particles correlation (right).
    The “short-range” ($\left| \Delta\eta \right| < 1.2$) and the normalized “long-range” ($1.2 < \left| \Delta\eta \right| < 1.4$) projections are shown in green and red points, respectively. 
    }
    \label{fig:1DWgt2PCF}
\end{figure}

To compute the $\langle \sumptns \rangle$ from the jet-like correlations, the combined, corrected correlation function is projected onto the $\Delta\varphi$ axis within $\left| \Delta\eta \right| < 1.2$.
This ``short-range'' projection is assumed to contain the full jet signal and the contribution from the uncorrelated background (underlying event).
Similarly, the projection within $1.2 < \left| \Delta\eta \right| < 1.4$ is also obtained.
This ``long-range'' projection is dominated by the underlying event and therefore can be used to provide an estimation of the uncorrelated background in the ``short-range''.
For this purpose, the ``long-range'' projection needs to be normalized per unit of rapidity and be subtracted from the ``short-range'' projection.
Examples are shown in Fig.~\ref{fig:1DWgt2PCF}, where the ``short-range'' projection is shown in green points and the normalized ``long-range'' projection is drawn in the red.
Then, $\langle \sumptns \rangle$ can be extracted from the jet-like correlations as the integral of the near-side peak region.
In most cases, the integration is performed within the $\left(-0.5\pi, 0.5\pi\right)$ region in the $\Delta\varphi$ in order to include all jet contributions.
However, for $0.6~<~\pts~<1.0$ \GeVc for \kzero and $0.6~<~\pts~<0.8$ \GeVc for \lmb~(\almb), a broader near-side peak is observed and the integration range is therefore extended to $\left( -0.6\pi, 0.6\pi \right)$.

\section{Systematic uncertainties}
\label{sec:SystUncertainty}
The systematic uncertainties are evaluated by varying the selection criteria or other analysis parameters. In each \pts interval, the individual systematic uncertainties are considered uncorrelated, and the total systematic uncertainty is calculated as the square root of the sum of their squares. 
For each source, the difference $\Delta$ between the varied and nominal results is divided by the corresponding statistical uncertainty, $\sigma_\Delta~=~\sqrt{\left|{\sigma_{var}^{2}-\sigma_{nom}^{2}}\right|}$, which is computed taking into account the correlation between the nominal and varied results.
The symbols $\sigma_{var}$ and $\sigma_{nom}$ denote the statistical uncertainties associated with the varied and nominal results, respectively~\cite{Barlow:2002yb}.
If $N_{\sigma}~=~\Delta/\sigma_{\Delta}~<~2$, the varied result is considered statistically compatible with the nominal result, and no systematic uncertainty for this source is assigned.
Otherwise, the contribution is added to the total systematic uncertainty.
The summary of all systematic uncertainties is given in Table~\ref{table:SystematicUn}.

The uncertainty due to the primary vertex $z$-position selection is estimated by tightening the allowed range from $\abs{z_{\rm vtx}}~<10$ cm to $\abs{z_{\rm vtx}}<~7$ cm, resulting in a more uniform detector acceptance.
For \kzero, the systematic uncertainty is negligible compared to the statistical one, and for \lmb~(\almb), the uncertainty rises to 0.6\% in the highest \pt interval.

The uncertainty due to the \Vzero signal extraction is determined by varying the signal and sideband regions, which accounts for the effect of the combinatorial background subtraction. 
The definition of the signal region is varied between 3$\sigma$ and 7$\sigma$ centered at the mean of the Gaussian function, and the corresponding lower and upper bounds of the sideband regions are varied between 4$\sigma$--7$\sigma$ and 8$\sigma$--14$\sigma$, respectively. 
For this source, the uncertainty is less than 0.4\% for \kzero and \lmb~(\almb) in most \pt intervals.
For \lmb~(\almb), it goes up to 1.2\% in the highest \pt interval.

To assess the systematic uncertainty of the geometrical acceptance correction, the scale factor $\alpha$ for the mixed-event distribution is computed using the inverse of the bin content centered at $\Delta\varphi~=~\Delta\eta~=~0$, rather than the default procedure, which uses the average evaluated at  $\Delta\eta~=~0$.
The systematic uncertainty from this source is below 0.5\% for \kzero and \lmb~(\almb) in most \pt intervals, rising to 1.2\% in the highest \pt interval for \kzero.

The systematic uncertainty related to the primary charged track selection is estimated by using a tighter DCA requirement. Tracks need to satisfy a \pt-dependent selection $\abs{{\rm DCA}_{xy}}~<~0.0105~+~0.0350/p_{\rm T}^{1.1}$~cm in the $xy$-plane and \abs{{\rm DCA}_{z}} $~<~2$ cm in the $z$-direction. 
For this source, the uncertainty is less than 0.4\% for \kzero and \lmb~(\almb).

The systematic uncertainty associated with the choice of the short-range region peak and the control region used for the subtraction of the uncorrelated background is assessed by changing the boundaries of these regions. 
The following boundaries for the short-range peak region (control region) are used for the estimation, \abs{\Delta\eta} $<1.1$ $(1.1~<~ \left | \Delta\eta \right |~<~1.4)$, \abs{\Delta\eta} $<1.3$ $(1.3~<~ \left | \Delta\eta \right |~<~1.4)$, \abs{\Delta\eta} $<1.3$ $(1.3~<~ \left | \Delta\eta \right |~<~1.5)$, and \abs{\Delta\eta} $<1.4$ $(1.4~<~ \left | \Delta\eta \right |~<~1.5)$.
For this source, the systematic uncertainty is less than 2.5\% in most \pt intervals for \kzero and \lmb~(\almb).

By default, the $\Vzero$ candidates are required to have at least one decay track with \ITS information to reject the out-of-bunch pileup.
As a variation of this selection, the candidates are required to have at least one decay track with information in \ITS and \TOF.
For this source, the uncertainty is less than 1\% in most intervals for \kzero.
For \lmb~(\almb), the systematic uncertainty from this source is negligible compared to the statistical one.

To take into account the systematic uncertainty related to the residual contamination from in-bunch (IB) pileup, the full sample is divided into two sets of equal size but different $\mu$, the average number of proton-proton interactions per bunch crossing.
The average $\mu$ value for the full sample is 0.0178, and for the two sub-samples, the corresponding values are 0.0085 and 0.0271, respectively.
For this source, the uncertainty is less than 2\% for \kzero and for most \pt intervals for \lmb~(\almb).

The imperfect description of the ALICE detector in the MC simulation~\cite{ALICE:2023kzv} leads to systematic biases in the estimation of the tracking efficiency and therefore of the \sumptns.
The corresponding systematic uncertainty of \meanz is calculated by varying the tracking efficiency by $\pm~3\%$ ~\cite{ALICE:2023ama, ALICE:2019wqv} and is found to be less than 1.4\% for \kzero and \lmb~(\almb).

Strange baryon \lmb and \almb candidates can also originate from the decay of $\Sigma^{\pm}$, $\Xi^{-}$, $\Omega^{-}$, and their corresponding antiparticles.
For the systematic uncertainty related to this source, the $\meanz$ of primary- and feed-down \lmb~(\almb) are studied in simulations.
The study indicates that $\meanz$ is not highly sensitive to the origin of the baryon.
For this source, a 2\% uncertainty uncorrelated with \pt is assigned.

\begin{table}[h]
    \caption{Systematic uncertainties of \meanz for \kzero and \lmb~(\almb). The uncertainties vary within the indicated ranges depending on the \pts interval.}
    \label{table:SystematicUn}
    \centering
    \begin{tabular}{| c | >{\centering\arraybackslash}m{2.0cm} | >{\centering\arraybackslash}m{2.0cm} |}
    \hline
    \multirow{2}{*}{Source} & \multicolumn{2}{c|}{\rule{0pt}{2.5ex}Systematic uncertainties (\%)} \rule{0pt}{2.5ex}\\
    \cline{2-3}
           & \rule{0pt}{2.5ex} $\kzero$ & $\lmb~(\almb)$ \rule{0pt}{2.5ex}\\
    \hline
    \multicolumn{1}{|l|}{$z_{\rm vtx}$ selection} & negligible & 0.0 -- 0.6 \\[0.2ex]
    \multicolumn{1}{|l|}{\Vzero signal extraction} & 0.0 -- 0.4 & 0.1 -- 1.2 \\[0.2ex]
    \multicolumn{1}{|l|}{Mixed-event scale} & 0.0 -- 0.5 & 0.0 -- 1.2 \\[0.2ex]
    \multicolumn{1}{|l|}{Primary charged track selection} & 0.0 -- 0.1 & 0.0 -- 0.4 \\[0.2ex]
    \multicolumn{1}{|l|}{Uncorrelated background subtraction} & 0.3 -- 2.5 & 0.5 -- 2.1 \\[0.2ex]
    \multicolumn{1}{|l|}{Out-of-bunch pileup} & 0.0 -- 1.1 & negligible \\[0.2ex]
    \multicolumn{1}{|l|}{In-bunch pileup} & 0.0 -- 1.6 & 0.1 -- 2.4 \\[0.2ex]
    \multicolumn{1}{|l|}{Material budget} & 1.1 -- 1.3 & 0.7 -- 1.4\\[0.2ex]
    \multicolumn{1}{|l|}{Feed-down} & N/A & 2.0 \\
    \hline
    \rule{0pt}{2.5ex} Total & 1.5 -- 2.9 & 2.3 -- 4.0 \rule{0pt}{2.5ex}\\
    \hline
    \end{tabular}
\end{table}

\section{Results and discussion}
\begin{figure}[h]
    \begin{center}
    \includegraphics[width = 0.7\textwidth]{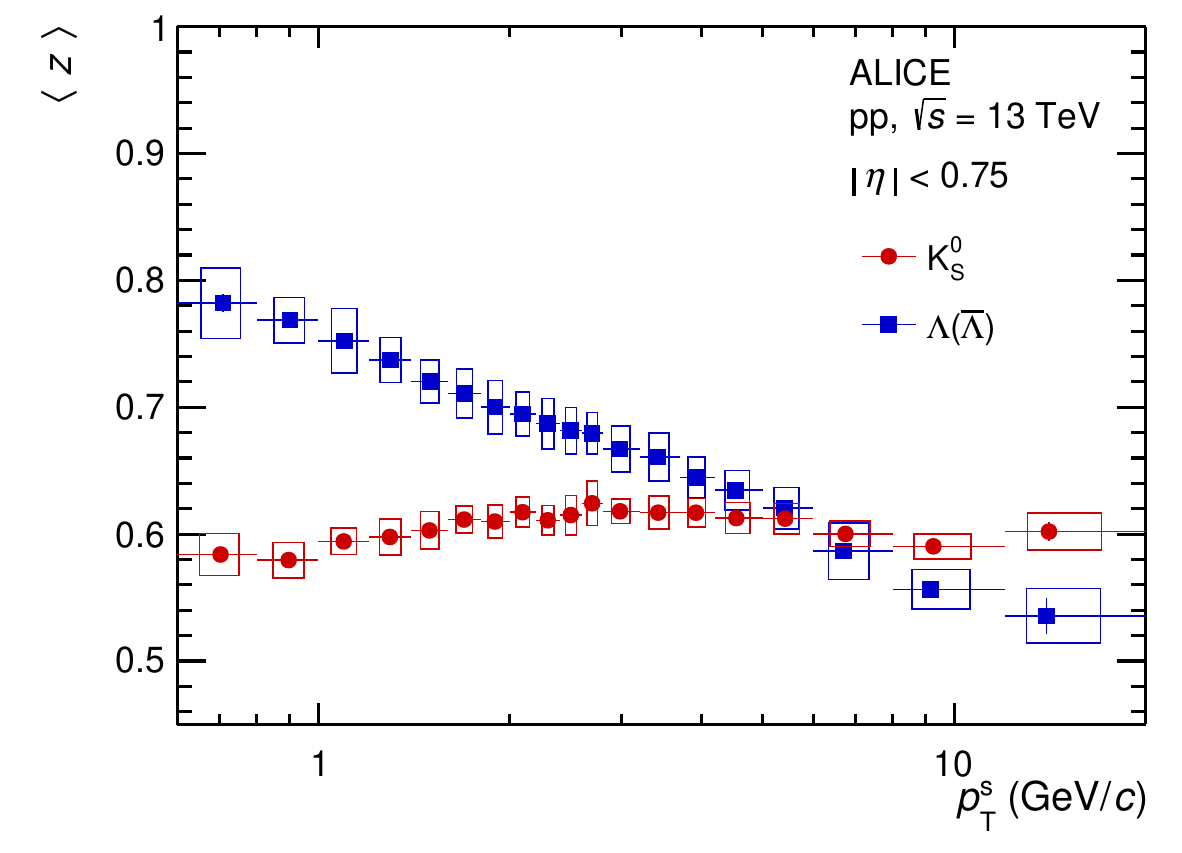}
    \end{center}
    \caption{The average transverse-momentum fraction (\meanz) for \lmb~(\almb) (blue) and \kzero (red) as a function of \pts in minimum bias \pp collisions at \thirteen. 
    The \pts ranges from 0.6 to 20 \GeVc.
    Statistical and systematic uncertainties of \meanz are represented by vertical error bars and empty boxes, respectively.
    Data points are drawn at the corresponding average \pts values in each interval.
    The horizontal bars represent the \pts interval widths.
    The statistical and systematic uncertainties of the average \pts values are negligible.}
    \label{fig:zInRD}
\end{figure}

The average transverse-momentum fractions \meanz as a function of strange particle \pts in \pp collisions at \thirteen for \kzero and \lmb produced in the $\etarange{0.75}$ are shown in Fig.~\ref{fig:zInRD}.
The summed transverse momentum of the associated particles ($\langle \sumptns \rangle$) has been measured for particles with $0.2 < \pt < 20$~\GeVc and no extrapolation outside this fiducial region has been performed.

At high \pts ($>6$~\GeVc), \meanz is approximately 0.6 for both particle species. This relatively large value compared to typical momentum fractions of light hadrons in jets (\meanz $\approx 0.3$)~\cite{ALICE:2023oww} is expected due to the trigger bias, characteristic of particles originating from the fragmentation of partons, whose \pt-differential production rate follows a steeply falling power-law spectrum (see discussion in Section~\ref{sec:PtWeighted2PC}).

For \kzero mesons, there is no indication of a significant change in the production mechanism over the measured \pts range as \meanz remains stable at $\approx 0.6$ within uncertainties.
On the other hand, interestingly, for the \lmb~(\almb) baryons, \meanz increases as \pts decreases, showing a rise from intermediate to low \pt and deviating from that of the \kzero below approximately 4~\GeVc.
For the lowest \pts interval, $\meanz = 0.78$, which is around 30\% higher than the high \pts limit.
Qualitatively, a rise can be expected from a steeper than power-law parent-particle \pt spectrum in the intermediate \pts region, which would worsen the aforementioned bias. 
However, if this were the only effect, it should be similar for \kzero and \lmb~(\almb). 
Hence, the increase of \meanz for \lmb~(\almb) is either because \lmb~(\almb) take a larger momentum fraction of the originating partons, or because there is a larger contribution from the isolated \lmb~(\almb) production in soft processes, which corresponds to $z~=~1$.
The observed difference in \meanz between \kzero and \lmb~(\almb) may also be attributed to strangeness and baryon number conservation.
For \lmb~(\almb), balancing can mainly occur via strange baryons, none of which contribute to \sumptns, while \kzero can be balanced by kaons, including the charged ones contributing to \sumptns.

\begin{figure}[h]
    \begin{center}
    \includegraphics[width = 0.7\textwidth]{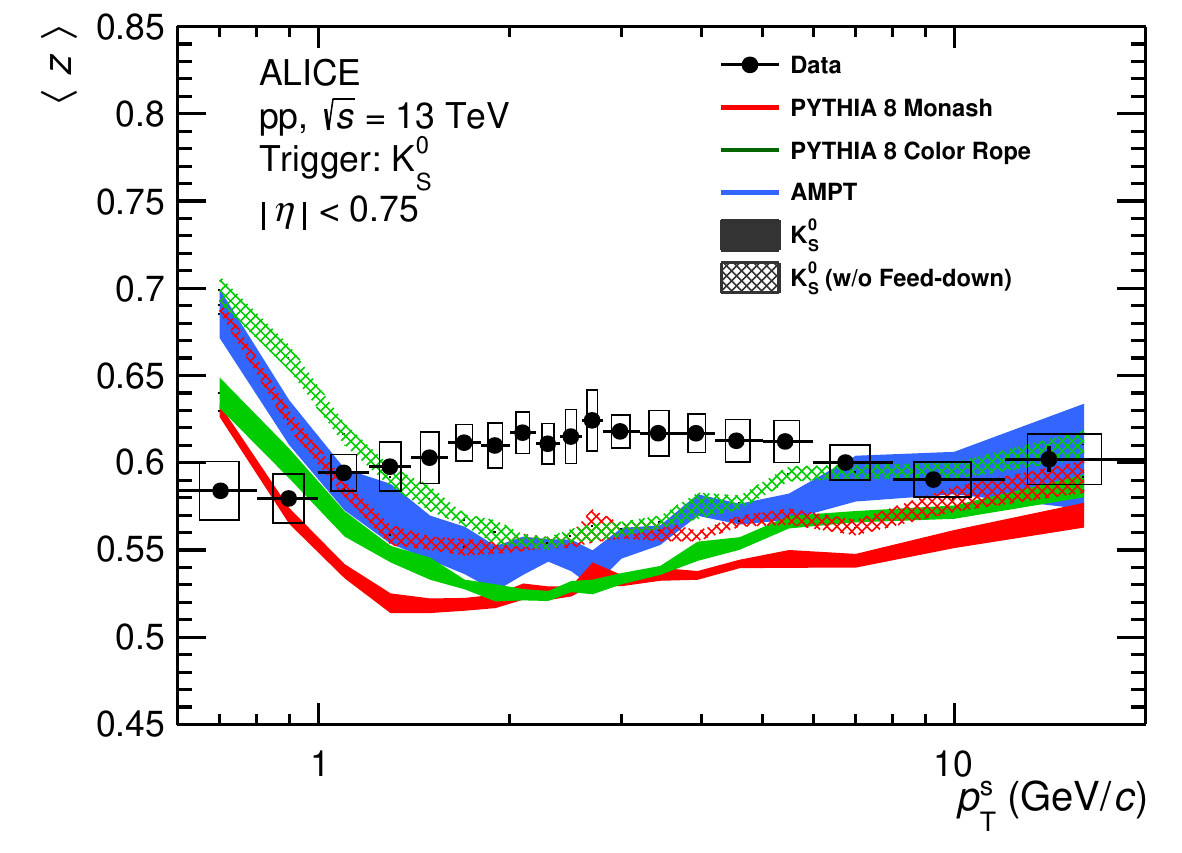}
    \includegraphics[width = 0.7\textwidth]{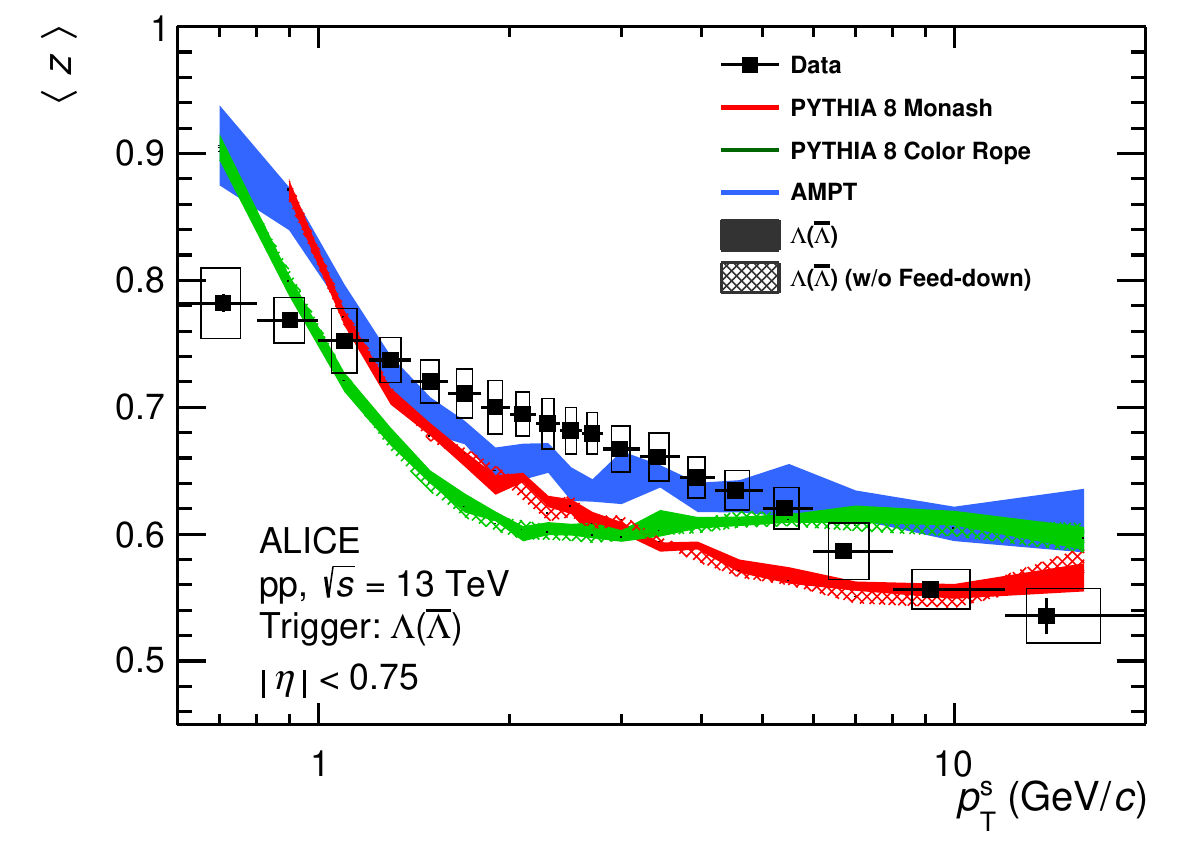}
    \end{center}
    \caption{The average transverse-momentum fraction (\meanz) for \kzero (top) and \lmb~(\almb) (bottom) in data compared with model calculations. The data are compared to \Pyeight with the Monash tune (red band), the Color Rope tune (green band) and the AMPT model with the string melting (blue band). For both \Pyeight tunes, the hatched bands represent the results excluding \kzero triggers coming from decays of the resonances ${\rm K}^{*}(892)^{0}$, ${\rm K}^{*}(892)^{\pm}$, and $\phi(1020)$, and \lmb~(\almb) triggers coming from decays of $\Sigma^{\pm}$, $\Xi^{-}$, $\Omega^{-}$ and their corresponding antiparticles.}
    \label{fig:zRDMC}
\end{figure}

In order to investigate potential differences in the hadronization mechanisms, the results are compared with several Monte Carlo model calculations employing different hadronization schemes, including \Pyeightthree~\cite{Bierlich:2022pfr} (Monash and Color Rope tunes) and AMPT (with string melting)~\cite{Lin:2004en}, as shown in Fig.~\ref{fig:zRDMC}.
The same kinematic and acceptance selections are applied in the simulation.
All primary charged particles are used as associated particles, excluding the long-lived charged strange baryons discussed in Section~\ref{sec:AnaApproach}, and the same analysis procedure is followed.
Unlike in data, the particle identity is known at generator level and no detector effects are involved.
Therefore, neither the side-band subtraction nor the additional weight $(1 - f_{\rm sec}) / \varepsilon$ is required for the correction on reconstruction inefficiency and secondary-particle contamination.
The band widths represent the quadratic sum of the statistical uncertainties and the systematic uncertainties associated with the uncorrelated background subtraction, as described in Section~\ref{sec:SystUncertainty}.
The hadronization mechanism in PYTHIA8 is based on the Lund string fragmentation model. 
Coherence between strings from different parton-parton interactions is implemented via the so-called color reconnection (CR) mechanism. 
Partons from different strings or parton showers are allowed to reconnect their color flow, potentially forming shorter or more energetically favorable strings before hadronization. 
The mechanism affects the final-state multiplicity and improves the description of the increase of the average \pt with multiplicity. 
Since version 8.2, Monash 2013 is the default tune.
It is a general-purpose tune that provides a consistent description of many observables measured at LHC and previous collider data~\cite{Skands:2014pea}.
Since CR alone is not able to describe strangeness enhancement, the Color Rope mechanism has been introduced.
Overlapping strings merge or interact forming ``ropes"---color flux tubes with higher color charge. 
These ropes have an increased effective string tension compared to normal strings which enhances strangeness production. 
In the AMPT model, the initial state is generated using HIJING~\cite{Wang:1991hta}, followed by partonic scatterings described by the Zhang’s Parton Cascade model~\cite{Zhang:1997ej}.
The hadronization process is then implemented through quark coalescence, where partons close in phase space are recombined to form hadrons.

As expected, in the high \pts region, the observed \meanz is well reproduced by the models.
In the intermediate \pts region, the models underestimate the data by approximately 10\%.
For \kzero, they predict a minimum close to 2 \GeVc and a steep rise towards lower \pts.
For \lmb~(\almb), the calculations show a steep rise below 2 \GeVc, without a significant minimum, unlike the case for \kzero.
In contrast, the measurement exhibits a smooth and continuous behavior over the measured \pts interval.

For the PYTHIA tunes, primary \Vzero production is compared to production including \kzero triggers originating from the decays of the resonances ${\rm K}^{*}(892)^{0}$, ${\rm K}^{*}(892)^{\pm}$, and $\phi(1020)$, and \lmb~(\almb) triggers coming from the decays of $\Sigma^{\pm}$, $\Xi^{-}$, $\Omega^{-}$ and their corresponding antiparticles, as shown in Fig.~\ref{fig:zRDMC} with solid-filled bands and hatched bands, respectively.
In the case of \lmb~(\almb), \meanz does not change significantly when excluding the feed-down contribution, while for \kzero, \meanz increases by at most about 10\%.

\section{Conclusions}
The first measurement of the average transverse-momentum fraction \meanz as a function of strange particle \pt for the strange baryon \lmb~(\almb) and the strange meson \kzero within mini-jets in \pp collisions at \thirteen is reported.
The results show different behavior for \kzero and \lmb~(\almb) in the low and intermediate \pt regions, suggesting different hadronization properties for mesons and baryons in these ranges.
The higher \meanz of \lmb~(\almb) observed in the intermediate \pt region could be attributed to either a harder fragmentation or a larger production of isolated \lmb.
This could be related to the enhancement of strange-baryon-to-meson yield ratios ($\Lambda / \kzero$) observed in previous measurements, in this \pt region~\cite{ALICE:2020jsh}. 
For \kzero, there is no strong \pt dependence of \meanz, which may indicate that there is no significant change of the production mechanism in the intermediate \pt region with respect to the high \pt region.
The \Pyeight Monash and Color Rope tunes, as well as the AMPT model, do not describe the \kzero data.
For \lmb~(\almb), the models overestimate the values at low \pt but reproduce the overall decreasing trend toward higher \pt, with AMPT showing the closest behavior in the intermediate region.
Overall, all used models fail to satisfactorily reproduce the measured \meanz.
They predict a different trend and strongly underestimate the \meanz of \kzero at the intermediate \pt, while for \lmb~(\almb), they show a steeper decrease than observed in the data at low and intermediate \pt.
Future studies will extend the analysis to multi-strange baryons and investigate the multiplicity dependence of the transverse-momentum fraction.
This will provide further information on the origin of the baryon enhancement in the intermediate \pts\ region.


\newenvironment{acknowledgement}{\relax}{\relax}
\begin{acknowledgement}
\section*{Acknowledgements}

The ALICE Collaboration would like to thank all its engineers and technicians for their invaluable contributions to the construction of the experiment and the CERN accelerator teams for the outstanding performance of the LHC complex.
The ALICE Collaboration gratefully acknowledges the resources and support provided by all Grid centres and the Worldwide LHC Computing Grid (WLCG) collaboration.
The ALICE Collaboration acknowledges the following funding agencies for their support in building and running the ALICE detector:
A. I. Alikhanyan National Science Laboratory (Yerevan Physics Institute) Foundation (ANSL), State Committee of Science and World Federation of Scientists (WFS), Armenia;
Austrian Academy of Sciences, Austrian Science Fund (FWF): [M 2467-N36] and Nationalstiftung f\"{u}r Forschung, Technologie und Entwicklung, Austria;
Ministry of Communications and High Technologies, National Nuclear Research Center, Azerbaijan;
Rede Nacional de Física de Altas Energias (Renafae), Financiadora de Estudos e Projetos (Finep), Funda\c{c}\~{a}o de Amparo \`{a} Pesquisa do Estado de S\~{a}o Paulo (FAPESP) and The Sao Paulo Research Foundation  (FAPESP), Brazil;
Bulgarian Ministry of Education and Science, within the National Roadmap for Research Infrastructures 2020-2027 (object CERN), Bulgaria;
Ministry of Education of China (MOEC) , Ministry of Science \& Technology of China (MSTC) and National Natural Science Foundation of China (NSFC), China;
Ministry of Science and Education and Croatian Science Foundation, Croatia;
Centro de Aplicaciones Tecnol\'{o}gicas y Desarrollo Nuclear (CEADEN), Cubaenerg\'{\i}a, Cuba;
Ministry of Education, Youth and Sports of the Czech Republic, Czech Republic;
The Danish Council for Independent Research | Natural Sciences, the VILLUM FONDEN and Danish National Research Foundation (DNRF), Denmark;
Helsinki Institute of Physics (HIP), Finland;
Commissariat \`{a} l'Energie Atomique (CEA) and Institut National de Physique Nucl\'{e}aire et de Physique des Particules (IN2P3) and Centre National de la Recherche Scientifique (CNRS), France;
Bundesministerium f\"{u}r Forschung, Technologie und Raumfahrt (BMFTR) and GSI Helmholtzzentrum f\"{u}r Schwerionenforschung GmbH, Germany;
National Research, Development and Innovation Office, Hungary;
Department of Atomic Energy Government of India (DAE), Department of Science and Technology, Government of India (DST), University Grants Commission, Government of India (UGC) and Council of Scientific and Industrial Research (CSIR), India;
National Research and Innovation Agency - BRIN, Indonesia;
Istituto Nazionale di Fisica Nucleare (INFN), Italy;
Japanese Ministry of Education, Culture, Sports, Science and Technology (MEXT) and Japan Society for the Promotion of Science (JSPS) KAKENHI, Japan;
Consejo Nacional de Ciencia (CONACYT) y Tecnolog\'{i}a, through Fondo de Cooperaci\'{o}n Internacional en Ciencia y Tecnolog\'{i}a (FONCICYT) and Direcci\'{o}n General de Asuntos del Personal Academico (DGAPA), Mexico;
Nederlandse Organisatie voor Wetenschappelijk Onderzoek (NWO), Netherlands;
The Research Council of Norway, Norway;
Pontificia Universidad Cat\'{o}lica del Per\'{u}, Peru;
Ministry of Science and Higher Education, National Science Centre and WUT ID-UB, Poland;
Korea Institute of Science and Technology Information and National Research Foundation of Korea (NRF), Republic of Korea;
Ministry of Education and Scientific Research, Institute of Atomic Physics, Ministry of Research and Innovation and Institute of Atomic Physics and Universitatea Nationala de Stiinta si Tehnologie Politehnica Bucuresti, Romania;
Ministerstvo skolstva, vyskumu, vyvoja a mladeze SR, Slovakia;
National Research Foundation of South Africa, South Africa;
Swedish Research Council (VR) and Knut \& Alice Wallenberg Foundation (KAW), Sweden;
European Organization for Nuclear Research, Switzerland;
Suranaree University of Technology (SUT), National Science and Technology Development Agency (NSTDA) and National Science, Research and Innovation Fund (NSRF via PMU-B B05F650021), Thailand;
Turkish Energy, Nuclear and Mineral Research Agency (TENMAK), Turkey;
National Academy of  Sciences of Ukraine, Ukraine;
Science and Technology Facilities Council (STFC), United Kingdom;
National Science Foundation of the United States of America (NSF) and United States Department of Energy, Office of Nuclear Physics (DOE NP), United States of America.
In addition, individual groups or members have received support from:
Czech Science Foundation (grant no. 23-07499S), Czech Republic;
FORTE project, reg.\ no.\ CZ.02.01.01/00/22\_008/0004632, Czech Republic, co-funded by the European Union, Czech Republic;
European Research Council (grant nos. 101220549, 950692), European Union;
Deutsche Forschungs Gemeinschaft (DFG, German Research Foundation) ``Neutrinos and Dark Matter in Astro- and Particle Physics'' (grant no. SFB 1258), Germany;
CONVECS project, CUP C97H23001700002 FESR 2021-2027 program, Italy.

\end{acknowledgement}

\bibliographystyle{utphys}   
\bibliography{bibliography}

\newpage
\appendix

%
%

\section{The ALICE Collaboration}
\label{app:collab}
\begin{flushleft} 
\small

D.A.H.~Abdallah\,\orcidlink{0000-0003-4768-2718}\,$^{\rm 134}$, 
I.J.~Abualrob\,\orcidlink{0009-0005-3519-5631}\,$^{\rm 112}$, 
S.~Acharya\,\orcidlink{0000-0002-9213-5329}\,$^{\rm 49}$, 
K.~Agarwal\,\orcidlink{0000-0001-5781-3393}\,$^{\rm II,}$$^{\rm 23}$, 
G.~Aglieri Rinella\,\orcidlink{0000-0002-9611-3696}\,$^{\rm 32}$, 
L.~Aglietta\,\orcidlink{0009-0003-0763-6802}\,$^{\rm 24}$, 
N.~Agrawal\,\orcidlink{0000-0003-0348-9836}\,$^{\rm 25}$, 
Z.~Ahammed\,\orcidlink{0000-0001-5241-7412}\,$^{\rm 132}$, 
S.~Ahmad\,\orcidlink{0000-0003-0497-5705}\,$^{\rm 15}$, 
I.~Ahuja\,\orcidlink{0000-0002-4417-1392}\,$^{\rm 36}$
% ...

\section*{Affiliation Notes}

$^{\rm I}$ Deceased\\
$^{\rm II}$ Also at: INFN Trieste, Trieste, Italy\\
$^{\rm III}$ Also at: Fondazione Bruno Kessler (FBK), Trento, Italy\\
$^{\rm IV}$ Also at: Czech Technical University in Prague, Prague, Czech Republic\\
$^{\rm V}$ Also at: Instituto de Fisica da Universidade de Sao Paulo\\
$^{\rm VI}$ Also at: Dipartimento DET del Politecnico di Torino, Turin, Italy\\
$^{\rm VII}$ Also at: University College of Dublin, Dublin, Ireland\\
$^{\rm VIII}$ Also at: Department of Applied Physics, Aligarh Muslim University, Aligarh, India\\
$^{\rm IX}$ Also at: Institute of Theoretical Physics, University of Wroclaw, Wroclaw, Poland\\
$^{\rm X}$ Also at: Facultad de Ciencias, Universidad Nacional Aut\'{o}noma de M\'{e}xico, Mexico City, Mexico\\

\section*{Collaboration Institutes}

$^{1}$ A.I. Alikhanyan National Science Laboratory (Yerevan Physics Institute) Foundation, Yerevan, Armenia\\
$^{2}$ AGH University of Krakow, Cracow, Poland\\
$^{3}$ Bogolyubov Institute for Theoretical Physics, National Academy of Sciences of Ukraine, Kyiv, Ukraine\\
$^{4}$ Bose Institute, Department of Physics  and Centre for Astroparticle Physics and Space Science (CAPSS), Kolkata, India\\
$^{5}$ California Polytechnic State University, San Luis Obispo, California, United States\\
$^{6}$ Central China Normal University, Wuhan, China\\
$^{7}$ Centro de Aplicaciones Tecnol\'{o}gicas y Desarrollo Nuclear (CEADEN), Havana, Cuba\\
$^{8}$ Centro de Investigaci\'{o}n y de Estudios Avanzados (CINVESTAV), Mexico City and M\'{e}rida, Mexico\\
$^{9}$ Chicago State University, Chicago, Illinois, United States\\
$^{10}$ China Nuclear Data Center, China Institute of Atomic Energy, Beijing, China\\
$^{11}$ China University of Geosciences, Wuhan, China\\
$^{12}$ Chungbuk National University, Cheongju, Republic of Korea\\
$^{13}$ Comenius University Bratislava, Faculty of Mathematics, Physics and Informatics, Bratislava, Slovak Republic\\
$^{14}$ Creighton University, Omaha, Nebraska, United States\\
$^{15}$ Department of Physics, Aligarh Muslim University, Aligarh, India\\
$^{16}$ Department of Physics, Pusan National University, Pusan, Republic of Korea\\
$^{17}$ Department of Physics, Sejong University, Seoul, Republic of Korea\\
$^{18}$ Department of Physics, University of California, Berkeley, California, United States\\
$^{19}$ Department of Physics, University of Oslo, Oslo, Norway\\
$^{20}$ Department of Physics and Technology, University of Bergen, Bergen, Norway\\
$^{21}$ Dipartimento di Fisica, Universit\`{a} di Pavia, Pavia, Italy\\
$^{22}$ Dipartimento di Fisica dell'Universit\`{a} and Sezione INFN, Cagliari, Italy\\
$^{23}$ Dipartimento di Fisica dell'Universit\`{a} and Sezione INFN, Trieste, Italy\\
$^{24}$ Dipartimento di Fisica dell'Universit\`{a} and Sezione INFN, Turin, Italy\\
$^{25}$ Dipartimento di Fisica e Astronomia dell'Universit\`{a} and Sezione INFN, Bologna, Italy\\
$^{26}$ Dipartimento di Fisica e Astronomia dell'Universit\`{a} and Sezione INFN, Catania, Italy\\
$^{27}$ Dipartimento di Fisica e Astronomia dell'Universit\`{a} and Sezione INFN, Padova, Italy\\
$^{28}$ Dipartimento di Fisica `E.R.~Caianiello' dell'Universit\`{a} and Gruppo Collegato INFN, Salerno, Italy\\
$^{29}$ Dipartimento DISAT del Politecnico and Sezione INFN, Turin, Italy\\
$^{30}$ Dipartimento di Scienze MIFT, Universit\`{a} di Messina, Messina, Italy\\
$^{31}$ Dipartimento Interateneo di Fisica `M.~Merlin' and Sezione INFN, Bari, Italy\\
$^{32}$ European Organization for Nuclear Research (CERN), Geneva, Switzerland\\
$^{33}$ Faculty of Electrical Engineering, Mechanical Engineering and Naval Architecture, University of Split, Split, Croatia\\
$^{34}$ Faculty of Nuclear Sciences and Physical Engineering, Czech Technical University in Prague, Prague, Czech Republic\\
$^{35}$ Faculty of Physics, Sofia University, Sofia, Bulgaria\\
$^{36}$ Faculty of Science, P.J.~\v{S}af\'{a}rik University, Ko\v{s}ice, Slovak Republic\\
$^{37}$ Faculty of Technology, Environmental and Social Sciences, Bergen, Norway\\
$^{38}$ Frankfurt Institute for Advanced Studies, Johann Wolfgang Goethe-Universit\"{a}t Frankfurt, Frankfurt, Germany\\
$^{39}$ Fudan University, Shanghai, China\\
$^{40}$ Gauhati University, Department of Physics, Guwahati, India\\
$^{41}$ Helmholtz-Institut f\"{u}r Strahlen- und Kernphysik, Rheinische Friedrich-Wilhelms-Universit\"{a}t Bonn, Bonn, Germany\\
$^{42}$ Helsinki Institute of Physics (HIP), Helsinki, Finland\\
$^{43}$ High Energy Physics Group,  Universidad Aut\'{o}noma de Puebla, Puebla, Mexico\\
$^{44}$ Horia Hulubei National Institute of Physics and Nuclear Engineering, Bucharest, Romania\\
$^{45}$ HUN-REN Wigner Research Centre for Physics, Budapest, Hungary\\
$^{46}$ Indian Institute of Technology Bombay (IIT), Mumbai, India\\
$^{47}$ Indian Institute of Technology Indore, Indore, India\\
$^{48}$ INFN, Laboratori Nazionali di Frascati, Frascati, Italy\\
$^{49}$ INFN, Sezione di Bari, Bari, Italy\\
$^{50}$ INFN, Sezione di Bologna, Bologna, Italy\\
$^{51}$ INFN, Sezione di Cagliari, Cagliari, Italy\\
$^{52}$ INFN, Sezione di Catania, Catania, Italy\\
$^{53}$ INFN, Sezione di Padova, Padova, Italy\\
$^{54}$ INFN, Sezione di Pavia, Pavia, Italy\\
$^{55}$ INFN, Sezione di Torino, Turin, Italy\\
$^{56}$ INFN, Sezione di Trieste, Trieste, Italy\\
$^{57}$ Inha University, Incheon, Republic of Korea\\
$^{58}$ Institute for Gravitational and Subatomic Physics (GRASP), Utrecht University/Nikhef, Utrecht, Netherlands\\
$^{59}$ Institute of Experimental Physics, Slovak Academy of Sciences, Ko\v{s}ice, Slovak Republic\\
$^{60}$ Institute of Physics, Homi Bhabha National Institute, Bhubaneswar, India\\
$^{61}$ Institute of Physics of the Czech Academy of Sciences, Prague, Czech Republic\\
$^{62}$ Institute of Space Science (ISS), Bucharest, Romania\\
$^{63}$ Institut f\"{u}r Kernphysik, Johann Wolfgang Goethe-Universit\"{a}t Frankfurt, Frankfurt, Germany\\
$^{64}$ Instituto de Ciencias Nucleares, Universidad Nacional Aut\'{o}noma de M\'{e}xico, Mexico City, Mexico\\
$^{65}$ Instituto de F\'{i}sica, Universidade Federal do Rio Grande do Sul (UFRGS), Porto Alegre, Brazil\\
$^{66}$ Instituto de F\'{\i}sica, Universidad Nacional Aut\'{o}noma de M\'{e}xico, Mexico City, Mexico\\
$^{67}$ iThemba LABS, National Research Foundation, Somerset West, South Africa\\
$^{68}$ Jeonbuk National University, Jeonju, Republic of Korea\\
$^{69}$ Korea Institute of Science and Technology Information, Daejeon, Republic of Korea\\
$^{70}$ Laboratoire de Physique Subatomique et de Cosmologie, Universit\'{e} Grenoble-Alpes, CNRS-IN2P3, Grenoble, France\\
$^{71}$ Lawrence Berkeley National Laboratory, Berkeley, California, United States\\
$^{72}$ Lund University Department of Physics, Division of Particle Physics, Lund, Sweden\\
$^{73}$ Marietta Blau Institute, Vienna, Austria\\
$^{74}$ Nagasaki Institute of Applied Science, Nagasaki, Japan\\
$^{75}$ Nara Women{'}s University (NWU), Nara, Japan\\
$^{76}$ National Centre for Nuclear Research, Warsaw, Poland\\
$^{77}$ National Institute of Science Education and Research, Homi Bhabha National Institute, Jatni, India\\
$^{78}$ National Nuclear Research Center, Baku, Azerbaijan\\
$^{79}$ National Research and Innovation Agency - BRIN, Jakarta, Indonesia\\
$^{80}$ Niels Bohr Institute, University of Copenhagen, Copenhagen, Denmark\\
$^{81}$ Nikhef, National institute for subatomic physics, Amsterdam, Netherlands\\
$^{82}$ Nuclear Physics Group, STFC Daresbury Laboratory, Daresbury, United Kingdom\\
$^{83}$ Nuclear Physics Institute of the Czech Academy of Sciences, Husinec-\v{R}e\v{z}, Czech Republic\\
$^{84}$ Oak Ridge National Laboratory, Oak Ridge, Tennessee, United States\\
$^{85}$ Ohio State University, Columbus, Ohio, United States\\
$^{86}$ Physics department, Faculty of science, University of Zagreb, Zagreb, Croatia\\
$^{87}$ Physics Department, Panjab University, Chandigarh, India\\
$^{88}$ Physics Department, University of Jammu, Jammu, India\\
$^{89}$ Physics Program and International Institute for Sustainability with Knotted Chiral Meta Matter (WPI-SKCM$^{2}$), Hiroshima University, Hiroshima, Japan\\
$^{90}$ Physikalisches Institut, Eberhard-Karls-Universit\"{a}t T\"{u}bingen, T\"{u}bingen, Germany\\
$^{91}$ Physikalisches Institut, Ruprecht-Karls-Universit\"{a}t Heidelberg, Heidelberg, Germany\\
$^{92}$ Physik Department, Technische Universit\"{a}t M\"{u}nchen, Munich, Germany\\
$^{93}$ Politecnico di Bari and Sezione INFN, Bari, Italy\\
$^{94}$ Research Division and ExtreMe Matter Institute EMMI, GSI Helmholtzzentrum f\"ur Schwerionenforschung GmbH, Darmstadt, Germany\\
$^{95}$ Saga University, Saga, Japan\\
$^{96}$ Saha Institute of Nuclear Physics, Homi Bhabha National Institute, Kolkata, India\\
$^{97}$ School of Physics and Astronomy, University of Birmingham, Birmingham, United Kingdom\\
$^{98}$ Secci\'{o}n F\'{\i}sica, Departamento de Ciencias, Pontificia Universidad Cat\'{o}lica del Per\'{u}, Lima, Peru\\
$^{99}$ SUBATECH, IMT Atlantique, Nantes Universit\'{e}, CNRS-IN2P3, Nantes, France\\
$^{100}$ Sungkyunkwan University, Suwon City, Republic of Korea\\
$^{101}$ Suranaree University of Technology, Nakhon Ratchasima, Thailand\\
$^{102}$ Technical University of Ko\v{s}ice, Ko\v{s}ice, Slovak Republic\\
$^{103}$ The Henryk Niewodniczanski Institute of Nuclear Physics, Polish Academy of Sciences, Cracow, Poland\\
$^{104}$ The University of Texas at Austin, Austin, Texas, United States\\
$^{105}$ Universidad Aut\'{o}noma de Sinaloa, Culiac\'{a}n, Mexico\\
$^{106}$ Universidade de S\~{a}o Paulo (USP), S\~{a}o Paulo, Brazil\\
$^{107}$ Universidade Estadual de Campinas (UNICAMP), Campinas, Brazil\\
$^{108}$ Universidade Federal do ABC, Santo Andre, Brazil\\
$^{109}$ Universitatea Nationala de Stiinta si Tehnologie Politehnica Bucuresti, Bucharest, Romania\\
$^{110}$ University of Cape Town, Cape Town, South Africa\\
$^{111}$ University of Derby, Derby, United Kingdom\\
$^{112}$ University of Houston, Houston, Texas, United States\\
$^{113}$ University of Jyv\"{a}skyl\"{a}, Jyv\"{a}skyl\"{a}, Finland\\
$^{114}$ University of Kansas, Lawrence, Kansas, United States\\
$^{115}$ University of Liverpool, Liverpool, United Kingdom\\
$^{116}$ University of Science and Technology of China, Hefei, China\\
$^{117}$ University of Silesia in Katowice, Katowice, Poland\\
$^{118}$ University of South-Eastern Norway, Kongsberg, Norway\\
$^{119}$ University of Tennessee, Knoxville, Tennessee, United States\\
$^{120}$ University of the Witwatersrand, Johannesburg, South Africa\\
$^{121}$ University of Tokyo, Tokyo, Japan\\
$^{122}$ University of Tsukuba, Tsukuba, Japan\\
$^{123}$ Universit\"{a}t M\"{u}nster, Institut f\"{u}r Kernphysik, M\"{u}nster, Germany\\
$^{124}$ Universit\'{e} Clermont Auvergne, CNRS/IN2P3, LPC, Clermont-Ferrand, France\\
$^{125}$ Universit\'{e} de Lyon, CNRS/IN2P3, Institut de Physique des 2 Infinis de Lyon, Lyon, France\\
$^{126}$ Universit\'{e} de Strasbourg, CNRS, IPHC UMR 7178, F-67000 Strasbourg, France, Strasbourg, France\\
$^{127}$ Universit\'{e} Paris-Saclay, Centre d'Etudes de Saclay (CEA), IRFU, D\'{e}partment de Physique Nucl\'{e}aire (DPhN), Saclay, France\\
$^{128}$ Universit\'{e}  Paris-Saclay, CNRS/IN2P3, IJCLab, Orsay, France\\
$^{129}$ Universit\`{a} degli Studi di Foggia, Foggia, Italy\\
$^{130}$ Universit\`{a} del Piemonte Orientale, Vercelli, Italy\\
$^{131}$ Universit\`{a} di Brescia, Brescia, Italy\\
$^{132}$ Variable Energy Cyclotron Centre, Homi Bhabha National Institute, Kolkata, India\\
$^{133}$ Warsaw University of Technology, Warsaw, Poland\\
$^{134}$ Wayne State University, Detroit, Michigan, United States\\
$^{135}$ Yale University, New Haven, Connecticut, United States\\
$^{136}$ Yildiz Technical University, Istanbul, Turkey\\
$^{137}$ Yonsei University, Seoul, Republic of Korea\\
$^{138}$ Affiliated with an institute formerly covered by a cooperation agreement with CERN\\
$^{139}$ Affiliated with an international laboratory covered by a cooperation agreement with CERN.\\

\end{flushleft} 

\end{document}